\documentclass[doublecol]{epl2}
\usepackage{amsmath}
\usepackage{amssymb}

\title{Double Entropic Stochastic Resonance}
\shorttitle{Double ESR}

\author{P.S. Burada\inst{1}
  \thanks{E-mail: {\tt Sekhar.Burada@physik.uni-augsburg.de}},
  G. Schmid\inst{1} 
  \thanks{E-mail: {\tt Gerhard.Schmid@physik.uni-augsburg.de}},
  D. Reguera\inst{2}, J.M. Rubi\inst{2},
  \and P. H\"anggi\inst{1}}
\shortauthor{P.S. Burada\etal}

\institute{
  \inst{1} Institut f\"ur Physik, Universit\"at Augsburg,
  Universit\"atsstr. 1, D-86135 Augsburg, Germany \\
  \inst{2} Departament de F\'isica Fonamental, Facultat de F\'isica,
  Universidad de Barcelona, Mart\'i i Franqu\'es 1, E-08028 Barcelona, Spain}

\pacs{02.50.Ey}{Stochastic processes}
\pacs{05.40.-a}{Brownian motion}
\pacs{05.10.Gg}{Stochastic analysis methods}

\date{Received: date / Revised version: date}

\abstract{We demonstrate the appearance of a purely entropic
stochastic resonance (ESR) occurring in a geometrically confined system,
where the irregular boundaries cause entropic barriers.
The interplay between a periodic input signal, a constant bias 
and intrinsic thermal noise leads to a resonant ESR-phenomenon 
in which feeble signals become amplified.
This new phenomenon is characterized by the presence of two peaks in the spectral 
amplification at corresponding optimal values of the noise strength. 
The main peak is associated with the manifest stochastic
resonance synchronization mechanism involving the inter-well
noise-activated dynamics while a second peak relates to a regime of
optimal sensitivity for intra-well dynamics. 
The nature of  ESR, occurring when the origin of the barrier is 
{\it entropic} rather than energetic, offers new perspectives for novel 
investigations and potential applications. ESR by itself presents yet 
another case where one constructively can harvest noise in driven 
nonequilibrium systems.}

\begin{document}

\maketitle

Stochastic resonance (SR) is an intriguing phenomenon occurring  in
systems pertaining to the wide class of periodically modulated noisy
systems which includes many different theoretical and experimental
situations \cite{gammaitoni,PT_SR,chemphyschem,
EPJ,vilar_mono1,vilar_mono2,Lutz99, Schmid01,Yasuda08,Murali}. 
SR relates to a remarkable idea that changed our common perception of noise
\cite{EPJ}: in particular, ambient noise may play a constructive
role in amplifying feeble signals or  may facilitate noisy
transport. The phenomenon has been well studied mainly in systems
having an intrinsic energetic potential whose origin is the presence
of interactions. However, the prevalent role of interactions in the
dynamics of a system is by no means general. It is indeed the free
energy what controls the dynamics and it could happen that the
entropic contribution plays a leading role. 
One example is the case of ion channels, where the role of entropic 
contributions for the phenomenon of SR has also been addressed
\cite{GoychukSR,GoychukSRPRE,GoychukSR1,Sung}.

Here, our focus is on stochastic resonance phenomena in confined systems,
where stylized, purely geometrical constrictions may lead to a
dominant entropic potential with a strong impact on the transport
characteristics \cite{Reguera_PRL, Burada_PRE, Burada_BioSy,
Burada_CPC} that may  exhibit a SR behavior in
some situations \cite{Burada_PRL, Burada_EPJB, Borromeo}. 
In this Letter we discuss a purely entropic resonant 
behavior in a geometrically
confined system in the presence of a longitudinal, constant bias and
an oscillating force. This new resonant phenomenon is distinctly
different and goes beyond the standard SR picture in several
aspects. First, its origin is strictly entropic and is solely
associated with geometric unevenness and confinement; in particular,
there is no energetic barrier in the system. Second, the situation
is characterized by the presence of two peaks in the amplification
factor signaling two different optimal values of the noise strength.
Finally, the enhancement of the amplification associated to the
second peak is not due to noise activation, but rather is due to an
optimal sensitivity of the intra-well dynamics to noise; it emerges
when the entropic barrier disappears.

\begin{figure}[t]
  \centering
  \onefigure{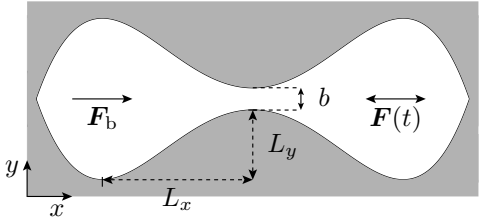}
  \caption{Schematic illustration of the two-dimensional
    structure confining the motion of the Brownian particles.
    The symmetric structure is defined by a quartic double well function,
    cf. eq.~(\ref{eq:widthfunctions}), involving the geometrical
    parameters $L_{x}$, $L_{y}$ and $b$.
    Brownian particles are driven by a sinusoidal
    force $F(t)$  and a constant
    bias $F_\mathrm{b}$ acting along the $x-$direction.}
\label{fig:well}
\end{figure}

To illustrate this phenomenon, we consider the dynamics of a Brownian
particle in the two-dimensional structure depicted in
fig.~\ref{fig:well}. The time evolution of this particle,
occurring in a constrained geometry subjected to a
sinusoidal oscillating force $F(t)$ and a constant bias
$F_\mathrm{b}$ acting along the longitudinal direction of the structure, 
can be described by means of the Langevin equation, written in the
overdamped limit as

\begin{align}
  \label{eq:langevin}
  \gamma \, \frac{\mathrm{d} \bm{r}}{\mathrm{d} t} = F_\mathrm{b} \,\bm{e_x}
  - F(t)\,\bm{e_x} +\sqrt{\gamma \, k_{\mathrm{B}}T}\, \bm{\xi}(t)\, ,
\end{align}
where $\bm{r}$ denotes the position of the particle, $\gamma$ is the
friction coefficient, $\bm{e_x}$ is the unit vector along the axial direction $x$,  and
$\bm{\xi}(t)$ is a Gaussian white noise with zero mean which
obeys the fluctuation-dissipation relation
$\langle \xi_{i}(t)\,\xi_{j}(t') \rangle = 2\, \delta_{ij}\,
\delta(t - t')$ for $i,j = x,y$.
The explicit form of the periodic input signal is given by
$F(t) = A \sin( \Omega t )$, where $A$ is the amplitude and $\Omega$
is the frequency of the sinusoidal signal.
In the presence of confining boundaries, this equation has to be solved by
imposing no-flow boundary conditions at the walls of the structure.

The shape of the considered 2D structure, fig.~\ref{fig:well}, which
is mirror symmetric about its $x-$ and $y-$ axis, is defined by
\begin{align}
  \label{eq:widthfunctions}
  w_{\mathrm{l}}(x) &= L_y \left(\frac{x}{L_x}\right)^4 -
  2\,L_y\left( \frac{x}{L_x}\right)^2 - \frac{b}{2} \, = -  w_{\mathrm{u}}(x)\, ,
\end{align}
where $ w_{\mathrm{l}}$ and $ w_{\mathrm{u}}$ correspond to the
lower and upper boundary functions, respectively, $L_{x}$ corresponds to
the distance between the location of the bottleneck and that of
maximal width, $L_{y}$ quantifies the narrowing of the boundary
functions and $b$ is the remaining width at the bottleneck, cf.
fig.~\ref{fig:well}. The local width of the structure is given by
$2\,w(x) = w_{\mathrm {u}}(x) - w_{\mathrm{l}}(x)$. This particular
choice of the geometry is intended to resemble the classical setup
for SR, namely a double well potential, where two basins are
separated by a barrier. However, we shall show that the distinct
nature of this confined system, dominated by a purely entropic
rather than energetic potential landscape, gives rise to new
phenomena.

For the sake of simplicity, we use reduced, dimensionless units. In
particular, we scale lengths by the characteristic length $L_x$,
time by $\tau=\gamma {L_x}^2/k_{\mathrm{B}}T_{\mathrm{R}}$, which is
the characteristic diffusion time at an arbitrary, although typical
reference temperature $T_{\mathrm{R}}$, force by
$F_{\mathrm{R}}=k_{\mathrm{B}}T_{\mathrm{R}}/L_x$, and frequency of
the sinusoidal driving by $1/\tau$. In dimensionless form the
Langevin equation ~\eqref{eq:langevin} and the boundary functions
\eqref{eq:widthfunctions} read:
\begin{align}
  \label{eq:dllangevin}
  \frac{\mathrm{d}\bm{r}}{\mathrm{d} t} & = F_\mathrm{b} \,\bm{e_x}
  - F(t)\,\bm{e_x} +\sqrt{D}\, \bm{\xi}(t)\, ,\\
  \label{eq:dlboundaryfunctions}
  w_{\mathrm{l}}(x) & = -w_{\mathrm{u}}(x) = - w(x)
  = \epsilon x^4 - 2\epsilon x^2
  - b/2 \, ,
\end{align}
where $D = T/T_{\mathrm{R}}$ is the dimensionless temperature
and $\epsilon = L_y/L_x$ is the aspect ratio of the structure.

{\bf Reduction of dimensionality. -- } In the absence of a
time-dependent applied force, i.e., $F(t) = 0 $, and tilting force,
i.e., $F_\mathrm{b} = 0$ it has been elaborated previously
\cite{Reguera_PRL, Burada_PRL, Burada_EPJB, Burada_PRE, Burada_BioSy, 
Burada_CPC,Zwanzig} that the 2D Fokker-Planck dynamics
corresponding to the Langevin equation \eqref{eq:dllangevin} can be
reduced to an effective 1D Fokker-Planck equation, reading in
dimensionless form
\begin{align}
\label{eq:fj}
\frac{\partial P(x,t)}{\partial t} =
     \frac{\partial}{\partial x}\left\{D \frac{\partial P(x,t)}{\partial x} \, +
     V^{\prime}(x,D)\, P(x,t) \right\} \, ,
\end{align}
where the prime refers to the derivative with respect to $x$ and
the effective potential 
\cite{Reguera_PRL,Burada_EPJB}
\begin{align}
\label{eq:effpotential}
V(x, D) = -D\, \ln[2\,w(x)] \, .
\end{align}

This equation describes the motion of a Brownian particle in a
bistable potential $V(x,D)$ of purely entropic nature. It is
important to emphasize that this bistable potential was not present
in the 2D Langevin dynamics, but arises due to the geometric
restrictions associated to confinement. 
In general, the diffusion
coefficient will depend on the coordinate $x$ as well
\cite{Burada_PRE, Burada_BioSy,Reguera_PRE,Zwanzig,Berezhkovskii2007}, but 
in the case discussed here this correction does not significantly
improve the accuracy of the reduced kinetics.
Therefore we do not further consider this correction.
Notably, for vanishing
width $2\,w(x)$ at the two opposite corners of the structure in
fig.~\ref{fig:well} this entropic potential approaches infinity,
thus intrinsically accounting for a natural reflecting boundary.

\begin{figure}[t]
  \centering
  \onefigure{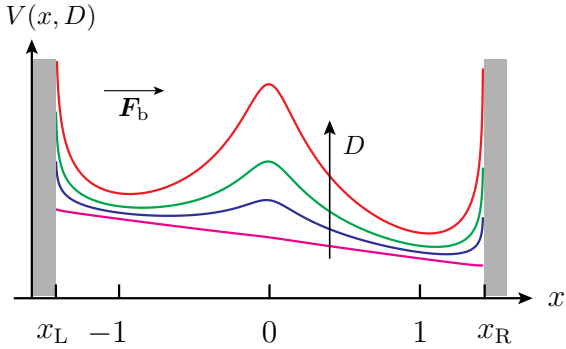}
  \caption{(Color online). The effective one-dimensional potential $V(x,D)$,
    eq.~(\ref{eq:free-tilt}), at a constant bias $F_\mathrm{b}$ and for 
    increasing noise strength $D$. The aspect ratio $\epsilon$ and the 
    bottleneck width $b$ of the structure are set to $1/4$ and $0.02$, respectively.
    The grey shaded regions on the left and right resemble the impenetrable
    boundaries. The potential function diverges at $x_\mathrm{L}$ and $x_\mathrm{R}$.}
\label{fig:pot-Ent}
\end{figure}

In the presence of a constant bias along the $x-$ direction
of the channel the free energy becomes 
\cite{Reguera_PRL,Burada_PRE, Burada_BioSy}
\begin{align}
\label{eq:free-tilt}
V(x,D) = -F_\mathrm{b}\, x - D\, \ln[2\,w(x)] \,.
\end{align}
The behavior of this tilted potential is depicted in
fig.~\ref{fig:pot-Ent}. In contrast to the purely energetic
potentials used for classical SR \cite{gammaitoni, hanggi}, the
barrier height and the shape of the potential change with noise
strength $D$. At small $D$ the strength of the entropic contribution
is almost negligible and the potential can be approximated by
$V(x,D) \approx -F_\mathrm{b}\, x$.
 The corresponding one-dimensional Fokker-Planck equation
 in the absence of periodic force reads
 \begin{align}
 \label{eq:reflecting}
 \frac{\partial P(x,t)}{\partial t} =
      \frac{\partial}{\partial x}\left\{D \frac{\partial P(x,t)}{\partial x} \, -
      F_\mathrm{b}\, P(x,t) \right\}\, .
 \end{align}

However, in general, the potential function $V(x,D)$,
eq.~(\ref{eq:free-tilt}), describes a tilted bistable potential,
where the height of the potential barrier depends on the noise
strength $D$. The location of the minima also depends on the noise
level, tending to a limiting value of $x = \pm 1$ for large $D$.
More importantly, the barrier disappears and the potential does not
exhibit any minima  when $D$ is smaller than a critical value
$D_{c}$, which depends on the geometrical structure and on the tilt.
For the shape given by eq.~(\ref{eq:widthfunctions}) with
$\epsilon=1/4$ and $b=0.02$, the potential has an inflection point
(i.e., the barrier disappears) when $D_c \approx 0.144 F_\mathrm{b}
$. We will  discuss below the relevance of this special situation in
the context of SR.

{\bf Spectral amplification. -- }
The response of the system to the weak periodic input signal, 
$F(t) = A\sin(\Omega\,t)$, is also a periodic function of time. 
The mean value of the position $x(t)$ in the asymptotic time limit
(i.e., after the memory of the
initial conditions is completely lost)
admits the Fourier series representation \cite{Jung89,Jung91}
\begin{align}
  \label{eq:fourierrepresentation}
  \langle x(t) \rangle_{as} = \sum_{n=-\infty}^{\infty}\, 
  M_{n}\,e^{i\,n\,\Omega t} \, , 
\end{align}
with the complex-valued amplitudes $M_{n}$ which depend nonlinearly on
the driving frequency, driving amplitude and on the noise strength.   

In order to study the occurrence of SR we analyzed the response of the system to the
applied sinusoidal signal in terms of the spectral amplification
$\eta$, which is defined as \cite{Jung89,Jung91}
\begin{align}
\label{eq:eta-Jung}
\eta = \left(\frac{2 |M_{1}|}{A} \right)^2\, .
\end{align}

{\bf Two-State approximation. -- }
If the potential barrier is sufficiently high and the two basins of attraction
are well separated, the intra-well motion may be neglected, and
a simplified two-state description can be used to get useful
insights into the full dynamics \cite{McNamara,Jung_ZP}.
Within the two-state approximation the transition rate from one well
to the other can be determined in terms of the
mean first passage time (MFPT) \cite{hanggi} to reach a potential
minimum starting out from the other minimum of the
tilted bistable potential $V(x,D)$. Accordingly, the forward ($k_{+}$) and
backward ($k_{-}$) rates are given by
\begin{align}
  \label{eq:rp}
  k_{\pm} = \frac{1}{T_{\pm}(\,x_{\mp}\, \to \,x_{\pm}\,)}\, ,
\end{align}
where $x_{-}$ and $x_{+}$ indicate the noise($D$)-dependent location of the
left and right minima, respectively;
$T_{+}(x_{-} \to x_{+})$ is the mean first passage time
for reaching $x_{+}$ starting out at $x_{-}$; and, vice-versa,
$T_{-}(x_{+} \to x_{-})$ is the mean first passage time in the backward direction.
More explicitly, the forward and backward mean first passage times are
\begin{align}
\label{eq:t1p}
T_{+} = \frac{1}{D} \int_{x_{-}}^{x_{+}} \mathrm{d}x\,
\frac{e^{-F_\mathrm{b}\,x/D}}{w(x)} \int_{x_\mathrm{L}}^{x} \mathrm{d}y \,
w(y)\,e^{F_\mathrm{b}\,y/D} \, ,
\end{align}
and
\begin{align}
\label{eq:t1m}
T_{-} = \frac{1}{D} \int_{x_{+}}^{x_{-}} \mathrm{d}x\, \frac{e^{-F_\mathrm{b}\,x/D}}{w(x)}
\int_{x_\mathrm{R}}^{x} \mathrm{d}y \,
w(y)\,e^{F_\mathrm{b}\,y/D} \, ,
\end{align}
where $x_\mathrm{L}$ and $x_\mathrm{R}$ are, respectively, the left and right limiting values
at which the boundary function vanishes, see fig.~\ref{fig:pot-Ent}.
Note that the bias $F_\mathrm{b}$ must be small 
in order to have two well separated basins of attraction
(see discussion of the potential).
Within the two-state description the mean position is given by
\begin{align}
  \label{eq:mean-ts}
  q_0 = \frac{k_{+}x_{+} + k_{-}x_{-}}{k_{+} + k_{-}} \,.
\end{align}

A sinusoidal signal leads to a modulation of the 
barrier height of the double well potential and consequently to a modulation 
of the rates $k_{\pm}$ \cite{McNamara}. 
One obtains the spectral amplification in the lowest oder of $A/D$, reading
\begin{align}
  \label{eq:eta-TS}
  \eta = \frac{1}{D^2} \left[\frac{(x_{+}-x_{-})^2 \,\,k_{+}\,k_{-}}
    {k_{+} + k_{-}} \right]^2 \, \frac{1}
  {(k_{+} + k_{-})^2 + {\Omega}^2} \,.
\end{align}
Note that for the symmetric case, i.e., $k_{+} = k_{-} = k$ and $x_{\pm} = \pm 1$,
the spectral amplification given in eq.~(\ref{eq:eta-TS}) reduces
to the well known expression,
\begin{align}
  \label{eq:eta-Sym}
  \eta = \frac{1}{D^2}\, \frac{4\,k^2}{4\,k^2 + {\Omega}^2} \,,
\end{align}
which in  the linear response regime is by construction independent
of the  amplitude strength \cite{gammaitoni,Jung_ZP,Jung89,Jung91}.

\begin{figure}[t]
  \onefigure{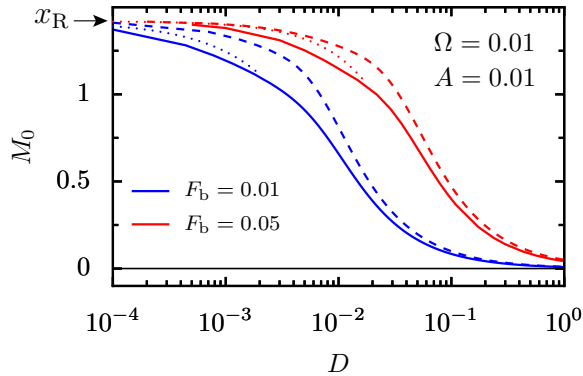}
  \centering
  \caption{(Color online). The time averaged mean position $M_0$ as a function of the
    noise strength $D$ for different tilting force values.
    The solid lines correspond to the numerical
    integration of the 1D Fokker-Planck equation, eq.~(\ref{eq:fj-1Dmodeling}),
    and the dashed lines correspond to the
    two-state approximation, eq.~(\ref{eq:mean-ts}), i.e.,  $M_0 \equiv q_0$.
    In the potential function we set the aspect ratio $\epsilon = 1/4$
    and the bottleneck width $b = 0.02$.
    The dotted lines represent the behavior of $M_0$ at small $D$
    obtained by integration of eq.~(\ref{eq:reflecting}).
    For the symmetric case ($F_\mathrm{b} = 0$) all these lines collapse
    to the  black solid line, i.e. $M_0=0$.}
  \label{fig:mean}
\end{figure}

{\bf 1D modeling. -- } The two-state description of the linear
response fails for either large driving amplitudes, small driving
frequencies\cite{gammaitoni,Casado_EPL} or very weak noise \cite{shneidman}. 
In such circumstances, the 1D
Fokker-Planck equation has to be considered and the nonlinear response
analyzed \cite{gammaitoni,Jung91,Casado_EPL,shneidman}.
In the presence of a sinusoidal signal $F(t)$ the 1D Fokker-Planck reads
\begin{align}
  \label{eq:fj-1Dmodeling}
  \frac{\partial P(x,t)}{\partial t} =
  \frac{\partial}{\partial x}\left\{D \frac{\partial P}{\partial x} \, +
   \left[V^{\prime}(x,D) + F(t) \right]\, P \right\} \, .
\end{align}

We have integrated numerically the above equation
by spatial discretization, using a Chebyshev collocation method, and
employing the method of lines to reduce the kinetic equation to a
system of ordinary differential equations, which was solved using a
backward differentiation formula method. Thereby we obtain
the time-dependent probability distribution $P(x,t)$ and
the time-dependent mean value of the position,
$\langle x(t) \rangle = \int x \, P(x,t)\, \mathrm{d} x $.
In the long-time limit this mean value approaches the periodicity of
the input signal with angular frequency $\Omega$ , see in
Refs.~\cite{Jung89,Jung91}. After a Fourier-expansion of $\langle
x(t) \rangle $ one finds the time averaged mean position
$M_0=M_{0}(A,D)$ and amplitude $M_1=M_{1}(A,D)$ of the first
harmonic of the response.

\begin{figure}[t]
  \onefigure{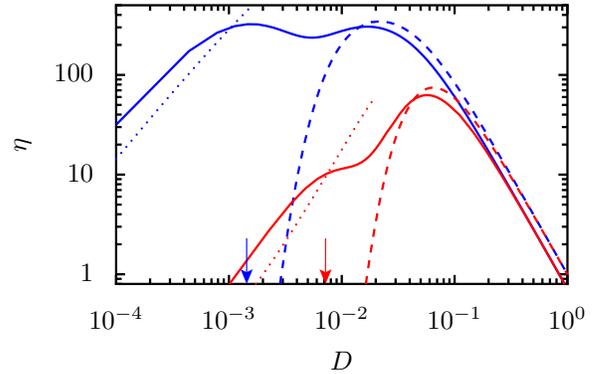}
  \centering
  \caption{(Color online). The behavior of the spectral amplification factor $\eta$
    as a function of the noise strength $D$
    for the same parameters as in fig.~\ref{fig:mean}.
    The solid lines correspond to the numerical integration of the
    1D Fokker-Planck equation, eq.~(\ref{eq:fj-1Dmodeling}), whereas the
    dashed lines correspond to the two-state approximation, eq.~(\ref{eq:eta-TS}).
    The behavior of $\eta$ at small $D$ obtained by integration of
    eq.~(\ref{eq:reflecting}) with a periodic forcing and at the respective
    tilting force values is represented with dotted lines.
    The arrows indicate the critical value $D_c$ that
    corresponds to the vanishing of the barrier. }
  \label{fig:eta-mfpt}
\end{figure}

The mean position $M_0$ as a function of the noise strength $D$ is
depicted in fig.~ \ref{fig:mean} for two different tilting force
values. For comparison, the approximated mean position $q_0$
obtained within the two-state modeling is also plotted. At very
small $D$  there is no barrier and the particle oscillates just in
the vicinity of the boundary, either at $x_\mathrm{R}$ or
$x_\mathrm{L}$ depending on the direction of the tilt.
As the noise strength increases the barrier height increases, see
fig.~\ref{fig:pot-Ent}, and the inter-well (from one well to the
other) dynamics become more dominant. Remarkably, the behavior of
the mean position can be qualitatively captured within the two-state
approximation.

For the 1D modeling the spectral amplification for the fundamental
oscillation is computed accordingly, cf. eq.~(\ref{eq:eta-Jung}).  
The comparison between the results of the 1D modeling and the
two-state approximation for the parameter $\eta$ as a function of
the noise strength $D$ and for two different tilting force values is
depicted in fig.~\ref{fig:eta-mfpt}. Within the 1D modeling $\eta$
exhibits a double-peak behavior for a finite tilt.
The appearance of the main peak at higher $D$, which is due to the
synchronization of the periodic signal with the activated inter-well
dynamics, can be nicely captured within the two-state model.
However, there is a second peak at small values of noise that cannot be
described within the two-state model. As it is discussed before, at
small noise strengths there is no barrier and the particle
oscillates in the vicinity of the boundary. As we increase the noise
one can observe a steep rise in $\eta$ with $D$ which is a
consequence of noise helping the particle to climb higher the
potential hill (see fig.~\ref{fig:eta-mfpt}). The second peak at small $D$
is attributed to the intra-well dynamics, and could also be
observed for mono-stable energetic potentials 
\cite{vilar_mono1,Reimann_EPL, Mayr}. The steepness of increase of $\eta$ depends on
the strength of the tilting force. However, beyond $D_c$, the
barrier appears and gets higher as we increase the noise level, thus
leading to a decrease in the amplification since the dynamics of the
particle is now hampered by the need to overcome a barrier.
Therefore, the inflection point of the effective potential -which
marks the appearance of the barrier- locates the position of a new
optimal regime of noise in terms of signal amplification.

{\bf 2D numerical simulation. -- }
In order to check the accuracy of the one-dimensional description
we compared the obtained results with the results of
Brownian dynamic simulations, performed by integrating the 
full overdamped Langevin equation \eqref{eq:dllangevin}.
The simulations were carried out using the standard stochastic Euler-algorithm.
\begin{figure}[t]
  \onefigure{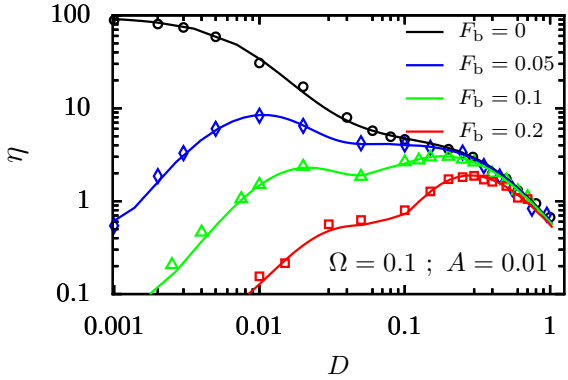}
  \centering
  \caption{(Color online). The spectral amplification $\eta$ 
    as a function of the noise strength $D$ at various tilting force values.
    The lines correspond to the numerical integration of the
    1D Fokker-Planck equation, eq.~(\ref{eq:fj-1Dmodeling}),
    whereas the symbols correspond to
    the results of the Langevin simulations for the two-dimensional
    structure with the shape defined by the dimensionless function
    $w(x) = -0.25\,x^4 + 0.5\,x^2 + 0.01$.}
  \label{fig:tilt}
\end{figure}

The behavior of the spectral amplification $\eta$ as a function
of the noise strength $D$ for different tilting force values
is depicted in fig.~\ref{fig:tilt}.
It is worth to mention that the results of the 1D modeling  (lines)
 are in very good agreement with the numerical simulations
of the full 2D-system (symbols) within a small relative error.
As one would expect the resonant behavior is absent
for zero tilting force, i.e., for the purely symmetric case
\cite{Burada_EPJB}, whereas in the presence of a tilting force, the spectral
amplification exhibits a double-peak structure.
As discussed before, the inter-well dynamics is responsible of the appearance
of the main peak at higher noise strengths $D$
whereas the second peak at smaller noise strengths is related to the disappearance of
the barrier. By changing the tilt, the second peak gets shifted to higher values of noise
following the predicted behavior for the inflection point,
namely $D_c \approx 0.144 F_\mathrm{b} $.

\begin{figure}[htb]
  \onefigure{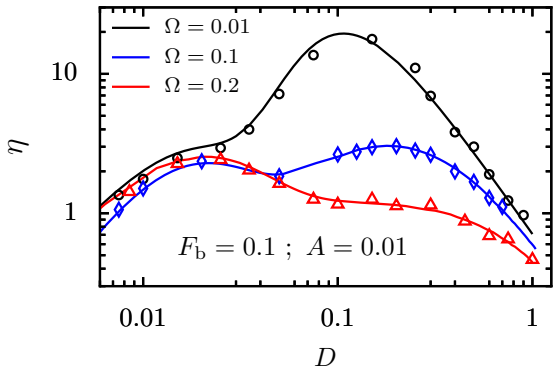}
  \centering
  \caption{(Color online).
    Same as in fig.~\ref{fig:tilt}, but for various 
    input signal frequencies.}
    \label{fig:SR-frequency}
  \end{figure}
  
The behavior of the spectral amplification for different frequencies
is plotted in fig.~\ref{fig:SR-frequency}. The height of the 
main peak in $\eta$ at high noise strength increases as the frequency of the input
signal decreases, resembling the behavior of classical SR \cite{PT_SR}. 
Overall, the perfect double-peak structure in the spectral amplification is
present only at moderate frequency range. A similar behavior of
double-peak SR one could also be observed in purely energetic 
systems, with a double-well potential, either at high 
input frequencies \cite{Jung91, Reimann_EPL} or at small frequencies 
in presence of  inertia effects \cite{Bulsara_PRE}.
\begin{figure}[t]
  \onefigure{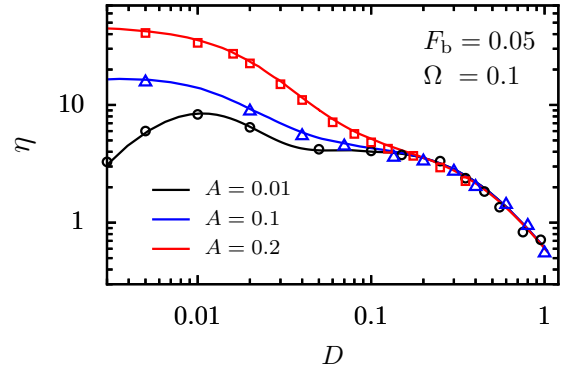}
  \centering
  \caption{(Color online).
    Same as in fig.~\ref{fig:tilt}, but for various input signal amplitudes.}
    \label{fig:SR-amplitude}
\end{figure}

Figure.~\ref{fig:SR-amplitude} depicts the behavior of the spectral
amplification for different values of the amplitude of the driving
force. The inter-well dynamics responsible of the appearance of the
main peak in $\eta$ at higher noise strengths is not much affected by the
variation of the input signal amplitude. In contrast, the second peak
is greatly affected by the amplitude of the input signal. A clear
double-peak structure in $\eta$ can be obtained only at small signal
amplitudes, since when the amplitude of the forcing exceed the bias
$F_\mathrm{b}$, the particle can move throughout the whole structure
assisted by the periodic forcing and therefore loses its sensitivity
to the noise strength.

{\bf Conclusions. -- } In this work we have studied the influence of
a constant bias on the resonant response of a geometrically
constrained system. In absence of the tilting force the system does
not exhibit a SR-behavior \cite{Burada_EPJB}. Applying a constant
bias, one can bring the system to an optimal regime where it
exhibits a double ESR in the spectral amplification. We want to
stress that the phenomenon is strictly entropic, since there is no
purely energetic barrier in the system. The ESR emerges due to the
interplay between the tilt and the entropic barrier which originates
in our case from the unevenness of geometrical restrictions. 
The double-peak behavior arises due to two different dynamic regimes:
(i) a regime, which is genuine of entropic systems, that occurs at high noise strengths  
where the main peak in the amplification is due to the synchronization of the 
periodic signal with noise-assisted hopping events; and 
(ii) a regime, at low noise strengths,
at which the potential has an inflection point.
The system thus becoming optimally sensitive to noise. 
Note that noise plays
a two-fold role in an entropic system: on the one hand, it facilitates
the dynamics, but on the other hand it intensifies the strength of
the entropic contribution leading to the appearance of barriers that
hamper transport. The second peak at smaller noise intensity signals
the optimum balance between these two roles.
This double resonant behavior renders the possibility for complete
control of the optimization of the response of a system in presence of noise. 
The main peak stemming from the inter-well dynamics can be strikingly controlled by 
tuning the frequency, while the second peak can be controlled either by 
changing the strength of the tilt or the amplitude 
of the periodic input signal. 
The double resonant peak behavior is a clear signature of the richness 
of dynamical behaviors in driven confined systems.
The occurrence of this phenomenon depends only on a few controllable 
parameters that can be advantageously tuned to achieve an optimal response.

\acknowledgments

This work has been supported by the DFG via the collaborative research
center, SFB-486, project A10, via the project no. 1517/26-2, the
Volkswagen Foundation (project I/80424), the German Excellence
Initiative via the \textit {Nanosystems Initiative Munich} (NIM), and
by the DGCyT of the Spanish government through grant No. FIS2008-04386.


\begin{thebibliography}{0}

\bibitem{gammaitoni}
  \Name{Gammaitoni L., H\"anggi P., Jung P., and Marchesoni F.}
  \REVIEW{Rev. Mod. Phys.}{70}{1998}{223}.

\bibitem{PT_SR}
  \Name{Bulsara A.R. and Gammaitoni L.}
  \REVIEW{Phys. Today}{49}{1996}{39}.

\bibitem{chemphyschem}
  \Name{H\"anggi P.}
  \REVIEW{ChemPhysChem}{3}{2002}{285}.

\bibitem{EPJ}
\Name{Gammaitoni L., H\"anggi P., Jung P., and Marchesoni F.}
\REVIEW{Eur. Phys. J.  B}{69}{2009}{1}.


\bibitem{vilar_mono1}
  \Name{Vilar J.M.G. and Rub\'i J.M.}
  \REVIEW{Phys. Rev. Lett.} {77}{1996}{2863}.


\bibitem{vilar_mono2}
  \Name{Vilar J.M.G. and Rub\'i J.M.}
  \REVIEW{Phys. Rev. Lett.} {78}{1997}{2886}.


\bibitem{Lutz99}
  \Name{Anishchenko V.S., Neiman A.B., Moss F., and Schimansky-Geier L.}
  \REVIEW{Phys. Usp.}{42}{1999}{7}.

\bibitem{Schmid01}
  \Name{Schmid G., Goychuk I., and H\"anggi P.}
  \REVIEW{Europhys. Lett.}{56}{2001}{22}.


\bibitem{Yasuda08}
  \Name{Yasuda H., Miyaoka T., Horiguchi J., Yasuda A.,H\"anggi P., and Yamamoto Y.}
  \REVIEW{Phys. Rev. Lett.} {100}{2008}{118103}.


\bibitem{Murali}
  \Name{Murali K., Sinha S., Ditto W.L., and Bulsara A.R.}
  \REVIEW{Phys. Rev. Lett.} {102}{2009}{104101}.


\bibitem{GoychukSR}
  \Name{Goychuk I. and H\"anggi P.}
  \REVIEW{Phys. Rev. Lett.}{91}{2003}{070601}.


\bibitem{GoychukSRPRE}
  \Name{Goychuk I. and H\"anggi P.}
  \REVIEW{Phys. Rev. E}{69}{2004}{021104}.


\bibitem{GoychukSR1}
  \Name{Goychuk I., H\"anggi P., Vega J.L., and Miret-Art\'es S.}
  \REVIEW{Phys. Rev. E}{71}{2005}{061906}.


\bibitem{Sung}
  \Name{Parc Y.W., Koh D., and Sung W.}
  \REVIEW{Eur. Phys. J. B}{69}{2009}{127}.


\bibitem{Reguera_PRL}
  \Name{Reguera D., Schmid G., Burada P.S., Rub\'i J.M., Reimann P., and H\"anggi P.}
  \REVIEW{Phys. Rev. Lett.}{96}{2006}{130603}.


\bibitem{Burada_PRE}
  \Name{Burada P.S., Schmid G., Reguera D., Rub\'i J.M., and H\"anggi P.}
  \REVIEW{Phys. Rev. E}{75}{2007}{051111}.


\bibitem{Burada_BioSy}
  \Name{Burada P.S., Schmid G., Talkner P., H\"anggi P., Reguera D., and Rub\'i J.M.}
  \REVIEW{BioSystems}{93}{2008}{16}.


\bibitem{Burada_CPC}
  \Name{Burada P.S., H\"anggi P., Marchesoni F., Schmid G., and Talkner P.}
  \REVIEW{ChemPhysChem}{10}{2009}{45}.


\bibitem{Burada_PRL}
  \Name{Burada P.S., Schmid G.,
    Reguera D., Vainstein M.H., Rubi J.M., and H\"anggi P.}
  \REVIEW{Phys. Rev. Lett.}{101}{2008}{130602}.


\bibitem{Burada_EPJB}
  \Name{Burada P.S., Schmid G.,
    Reguera D., Rubi J.M., and H\"anggi P.}
  \REVIEW{Eur. Phys. J. B}{69}{2009}{11}.

\bibitem{Borromeo}
  \Name{Borromeo M. and Marchesoni F.}
  \REVIEW{Eur. Phys. J. B}{69}{2009}{23}.


\bibitem{Zwanzig}
  \Name{Zwanzig R.}
  \REVIEW{J. Phys. Chem.}{96}{1992}{3926}.


\bibitem{Reguera_PRE}
  \Name{Reguera D. and Rub\'i J.M.}
  \REVIEW{Phys. Rev. E}{64}{2001}{061106}.


\bibitem{Berezhkovskii2007}
  \Name{Berezhkovskii A.M., Pustovoit M.A., and Bezrukov S.M.}
  \REVIEW{J. Chem. Phys.}{126}{2007}{134706}.


\bibitem{hanggi}
  \Name{H\"anggi P., Talkner P., and Borkovec M.}
  \REVIEW{Rev. Mod. Phys.}{62}{1990}{251}.


\bibitem{Jung89}
  \Name{Jung P. and H\"anggi P.}
  \REVIEW{Europhys. Lett.}{8}{1989}{505}.


\bibitem{Jung91}
  \Name{Jung P. and H\"anggi P.}
  \REVIEW{Phys. Rev. A}{44}{1991}{8032}.


\bibitem{McNamara}
  \Name{McNamara B. and Wiesenfeld K.}
  \REVIEW{Phys. Rev. A}{39}{1989}{4854}.


\bibitem{Jung_ZP}
  \Name{Jung P. and H\"anggi P.}
\REVIEW{Z. Physik B}{90}{1993}{255}.


\bibitem{Casado_EPL}
  \Name{Casado-Pascual J., G\'omez-Ord\'o\~nez J., Morillo M., and H\"anggi P.}
  \REVIEW{Europhys. Lett.}{65}{2004}{7}.

\bibitem{shneidman}
  \Name{Shneidman V.A., Jung P., and H\"anggi P.}
  \REVIEW{Phys. Rev. Lett. }{72}{1994}{2682}.


\bibitem{Reimann_EPL}
  \Name{Evstigneev M., Reimann P., Pankov V., and Prince R. H.}
  \REVIEW{Europhys. Lett.}{65}{2004}{7}.

\bibitem{Mayr}
  \Name{Mayr E., Schulz M., Reineker P., Pletl T., and Chvosta P.}
  \REVIEW{Phys. Rev. E}{76}{2007}{011125}.

\bibitem{Bulsara_PRE}
  \Name{Alfonsi L., Gammaitoni L., Santucci S., and Bulsara A.R.}
  \REVIEW{Phys. Rev. E}{62}{2000}{299}.


\end{thebibliography}
\end{document}